\documentclass[12pt]{iopart}

\usepackage{graphicx}
\begin{document}

\title{Time and Frequency Domain Polariton Interference}
\author{G. Campbell, M. Hosseini,  B. M. Sparkes, P. K. Lam, and B. C. Buchler}
\address{ Centre for Quantum Computation and Communication Technology, Department of Quantum Science, The Australian National University, Canberra, Australia}

\date{\today}

\begin{abstract} 
We present experimental observations of interference between an atomic spin coherence and an optical field in a $\Lambda$-type gradient echo memory. The interference is mediated by a strong classical field that couples a weak probe field to the atomic coherence through a resonant Raman transition. Interference can be observed between a prepared spin coherence and another propagating optical field, or between multiple $\Lambda$ transitions driving a single spin coherence. In principle, the interference in each scheme can yield a near unity visibility.
\end{abstract}

\maketitle

Coherent manipulation of atomic systems using photons is a key element of many quantum-atom optics experiments. The ability to controllably tune atom-light interactions while preserving the quantum properties of a system also has great potential with regard to the development of quantum information technology. Many of the techniques employed in quantum atom-optics involve interaction of light with ensembles of atoms that have long-lived coherences between hyperfine energy levels. In such systems, a two-photon transition between hyperfine states can be used to manipulate the atomic state in a coherent manner. Examples of this include stimulated Raman adiabatic passage (STIRAP)  \cite{RevModPhys.70.1003}, electromagnetically induced transparency (EIT) \cite{RevModPhys.77.633,Fleischhauer:2000p7451} and photon echoes \cite{Hartmann:echoint:OL:1993}, all of which have been proposed as central elements in a range of protocols for storing and processing optical quantum information.

\begin{figure}
\centerline{\includegraphics[width=0.8\columnwidth]{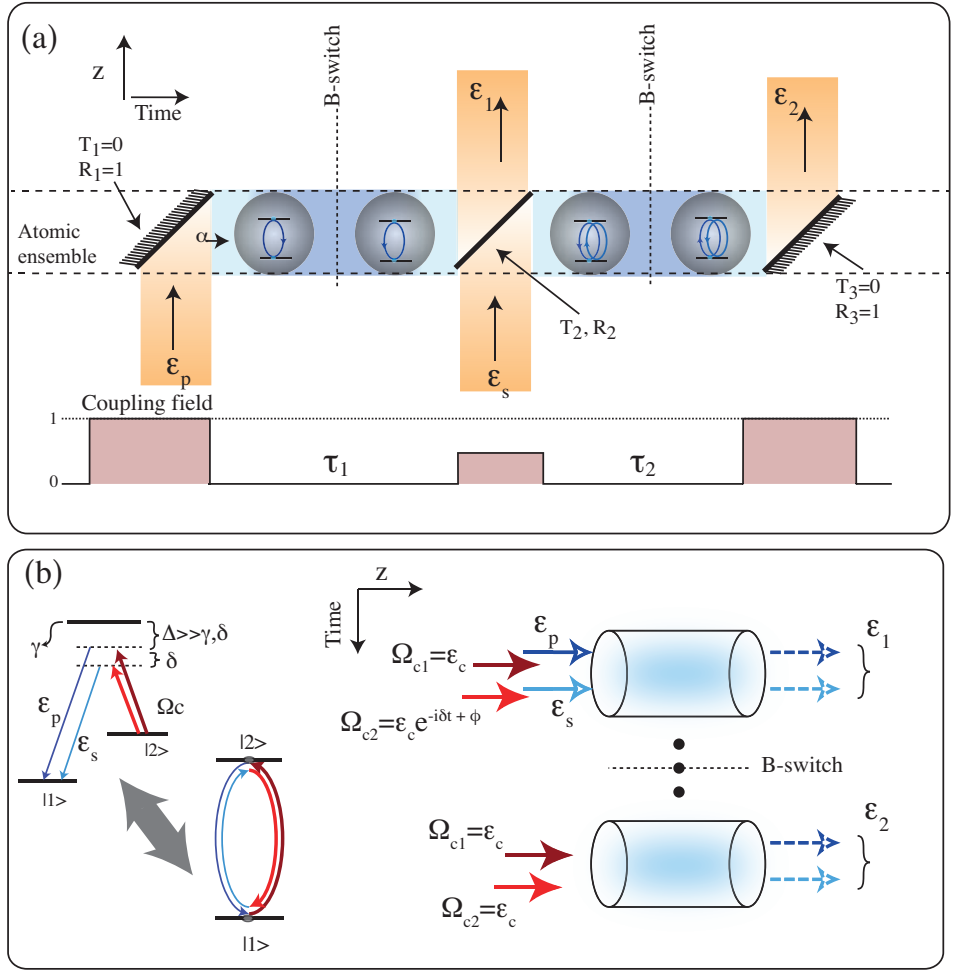}}
 \caption{ (a) Schematic representation of atom-light interference in the memory. The probe pulse, $\mathcal{E}_p$, is fully absorbed in the atomic spin coherence ($\alpha$). The second steering pulse, $\mathcal{E}_s$, enters the memory at the precise time that the first echo is being emitted so that it can interfere with the recalled light. The interference is determined by relative phase of the pulses and the effective beamsplitter ($T_2,R_2$), which is controlled by the strength of the Raman coupling field. The remaining atomic coherence can be recalled later as $\mathcal{E}_2$. (b) Left: Double-$\Lambda$ level structure and optical fields used for interference of two Raman absorption paths of signal fields (probe and steering) with different frequencies. Both $\Lambda$ transitions drive the same coherence. Right: The procedure used for observing double-Raman interference. The probe and steering pulses are sent into the memory each with a corresponding coupling field. Interference between the unabsorbed probe pulses, $\mathcal{E}_1$, and the atomic coherence, which is recalled from the memory as $\mathcal{E}_2$, can be observed by varying the relative phase of the two $\Lambda$ transitions.}
 \label{scheme}
\end{figure} 

Within the range of schemes that exploit light-atom interactions, a number of them, particularly photon echo schemes, pertain to interference effects between the quantum modes. A time-delayed quantum interferometer has previously been proposed as a method for quantum interference between two single photons \cite{Moiseev:TDSP:2001}. Experimental observation of interference between backward-propagating stimulated photon echoes has also been reported \cite{Hartmann:echoint:OL:1993}, where two echoes have been selectively chosen in time to destructively interfere while the information contained in the suppressed echo was not recovered from the sample. Furthermore, the phase preserving nature of storage was previously investigated  by interfering echoes generated from separate optical  memories \cite{Gisisn:IntEcho:PRL:2007}.

In this paper, we investigate the coherent interference of atomic spin coherence and an optical field using the three-level gradient echo memory ($\Lambda$-GEM) technique  \cite{Hetet:2008p5840,Hosseini:2009p8466,Hosseini:NComm:2011,Hosseini:Nphys:2011}. We treat the read and write stages of the memory as being analogous to a beam-splitting operation acting between an optical mode and an atomic coherence. Both the splitting ratio and the interference phase are controlled optically via the strength of the Raman coupling field.

We explore the nature of the atom-light coupling through two separate mechanisms. The first scheme (Fig.~\ref{scheme}(a)) is a time-domain interferometer. We prepare the atomic coherence by storing a pulse of light in the atomic memory. A second 'steering pulse' is sent into the memory just as the stored field is being recalled, leading to interference that supresses or enhances the recall of the stored pulse. The second scheme (Fig.~\ref{scheme}(b)) is a frequency domain interferometer. In this case the atomic coherence is simultaneously driven via two distinct Raman transitions. Interference occurs between the two nondegenerate absorption paths. Before examining these schemes in detail we will outline our coherent optical storage scheme.

The $\Lambda$-GEM technique operates by applying a spatially linearly varying detuning to an ensemble of three-level atoms. A strong coupling field is then used to couple a weak probe field to the long-lived atomic spin coherence, which dephases due to the applied detuning gradient. By reversing the sign of the gradient, the atomic dipoles can be rephased, resulting in a photon echo. It has been shown previously that $\Lambda$-GEM is capable of efficient and noiseless storage of quantum states of light \cite{Hosseini:NComm:2011,Hosseini:Nphys:2011}. Moreover, this scheme has the capacity to arbitrarily access stored bits of information \cite{Hosseini:2009p8466} and manipulated the stored information in the time and frequency domains \cite{Buchler:2010p11952}.

The behaviour of the GEM is best understood using a spatial Fourier decomposition (with $k$ being the spatial frequency) of the optical field $\hat{\mathcal{E}}(t,k)$ and the coherence, $\hat{\sigma}_{12}(t,k)$, between the atomic energy levels $\vert 1 \rangle$ and $\vert 2 \rangle$. The normal mode of the two-level GEM is defined as $\hat{\psi}(t,k)=k\hat{\mathcal{E}}(t,k) + N\hat{\sigma}_{12}(t,k)$ \cite{Hetet:2008p7696} and propagates in the $(t,k)$ plane where $N$ is the linear atomic density. Like the normal mode in EIT\cite{Fleischhauer:2000p7451}, $\hat{\psi}(t,k)$ is a combination of atomic polarisation and optical field and can be considered a polariton. The velocity at which the polariton propagates in $k$-space is proportional to the atomic frequency gradient and, by switching the sign of the magnetic field gradient, the evolution of the polariton can be reversed. When the polariton again reaches $k=0$ the atomic coherence is rephased and the polariton is recalled as an optical field.

A similar polariton equation, $\hat{\psi}(t,k)=k\hat{\mathcal{E}}(t,k) + \frac{ N \Omega}{\Delta} \hat{\sigma}_{12}(t,k)$, can be defined for the $\Lambda$-GEM system by replacing $N$ with $N\frac{\Omega_{c}}{\Delta}$, where $\Omega_c$ is the coupling field Rabi frequency and $\Delta$ is the Raman detuning from the excited state. In the three-level case $\sigma_{12}(t,k)$ describes the atomic spin coherence.

The polariton propagates in the $(t,k)$ plane following 
\begin{equation}
\left( \frac{\partial}{\partial t} \right.-\left. \eta(t)
\frac{\partial}{\partial k}-i\frac{g N \Omega_{c}^{2}}{k \Delta^{2}} \right)  \hat{\psi} (t,k)=0,
\label{motion}
\end{equation}
where $g$ is the atom-light coupling strength and $\eta$ is the detuning gradient that results from the applied magnetic field \cite{Hetet:2008p7696}. The evolution of the three-level GEM is similar to that of the two-level GEM but the addition of the coupling field results in additional flexibility: the strength and phase of the coupling between the probe field and the memory can be optically controlled. 

Our first experiment investigates interference of light pulses with a mode stored in the atomic memory. Following on from Ref. \cite{Longdell:2008p8530}, this effect can be thought of as a time-delayed beamsplitter system. The effective optical depth (OD) of a $\Lambda$-GEM is defined as   
\begin{eqnarray}
\beta=\frac{gN}{\eta}(\frac{\Omega_c}{\Delta})^2,
\end{eqnarray}
where $g$ is the coupling strength and $\eta$ is the linear broadening from the magnetic field.

For the writing stage the transmissivity, $T(\beta)$, of the effective beamsplitter is the fraction of the input light field that is leaked through the memory so that $T(\beta)=e^{- 2\pi \beta}$, while the fraction of the light written into the memory is given by the reflectivity $R(\beta)=1-T(\beta)$. For the reading stage, the $R(\beta)$ will be the fraction of the polariton that is converted into a recalled optical field while $T(\beta)$ will be the fraction that remains in the memory. Since $T(\beta)$ and $R(\beta)$ are defined by the strength of the coupling field, one can tune the transmittivity of the beam-splitting through the power of the coupling field. A series of reading and writing events, as shown in Fig.~\ref{scheme}(a), can then be described using appropriate reflectivites.  The amount of light recalled in the first echo is given by $\mathcal{E}_1=\sqrt{R_1 R_2} e^{-\gamma_0 \tau}\mathcal{E}_p + e^{i\theta}\sqrt{T_2} \mathcal{E}_s$, where $\mathcal{E}_p$ is an initial probe pulse and $\mathcal{E}_s$ is a second pulse, which we label the steering pulse, that enters the medium at the time that the probe pulse is recalled. The exponential term arises from the decay of $\mathcal{E}_p$, at a rate $\gamma_0$, during the storage time $\tau$ and the phase $\theta$ can be chosen at will.  This equation shows that interference can arise between recalled fraction of $\mathcal{E}_p$ and $\mathcal{E}_s$ and, in particular, if $\sqrt{R_1 R_2} e^{-\gamma_0 \tau}=\sqrt{T_2}$ and $\theta=\pi$ then $\mathcal{E}_1$ can be fully suppressed. This simple analysis ignores other details such as the matching of the temporal modes of the pulses. Other factors that limit ideal interference will be discussed later when we analyse the results of our experiments.

We use the polariton picture to visualise the dynamics of the time-domain beamsplitting operation. As the probe pulse is absorbed and the atomic coherence dephases, the polariton evolves from to higher spatial frequencies and becomes primarily an atomic spin-wave. The position of the mode in k-space represents how much the atoms in the spin wave have dephased.  A higher atomic frequency gradient means there is a larger dephasing and therefore faster progression to higher values of k. When the magnetic field gradient is reversed, the polariton propagates back towards $k=0$ at which point the coherence is rephased and the polariton is converted into an optical field which exits the memory. The $k=0$ crossing is analogous to the interface of a beamsplitter. At this point the $\Lambda$ transition couples the optical mode of the probe field, propagating in the $(t,z)$ plane, and the spin coherence of the atomic ensemble, propagating in the $(t,k)$ plane.
 
\begin{figure}[!h]
  \centerline{\includegraphics[width=0.6 \columnwidth]{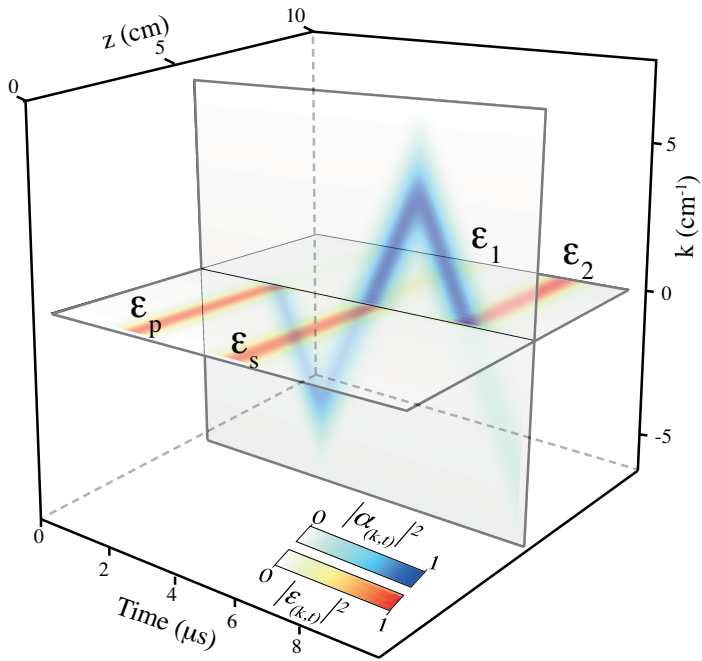}}
  \caption{Numerical simulation showing interference between the electric field, plotted on the z-t plane, and the atomic coherence, plotted on the k-t plane, where the second light pulse is out of phase from the first echo field. The parameters used in the simulations are: $gNL/\gamma=40$, $\Omega_c/\Delta=0.75$, $\Omega_c(t=4\mu s)=0.7\Omega_c(t=2\mu s)$ and $\phi_{\mathcal{E}_s}-\phi_{\mathcal{E}_p}=\pi$. The magnitudes of the electric field and the atomic coherence are plotted.}
  \label{3Dxmds}
\end{figure}  
 
Figure \ref{3Dxmds} shows a numerical simulation of the time-domain interference scheme using the methods described in  \cite{Hetet:2008p7696}. The simulation shows the evolution of the electric field in real space (the $(t,z)$ plane) and the atomic spin coherence in Fourier-space (the $(t,k)$ plane). In this numerical simulation, the phase of the steering pulse is chosen such that a suppression of the echo from the probe pulse is observed. The intensity of the coupling field is chosen so that the effective splitting ratio is roughly 0.5. Constructive interference into the atomic polarisation occurs and the resultant polariton is recalled in field $\mathcal{E}_2$ after the second gradient switch.

\begin{figure}
 \centerline{\includegraphics[width=\columnwidth]{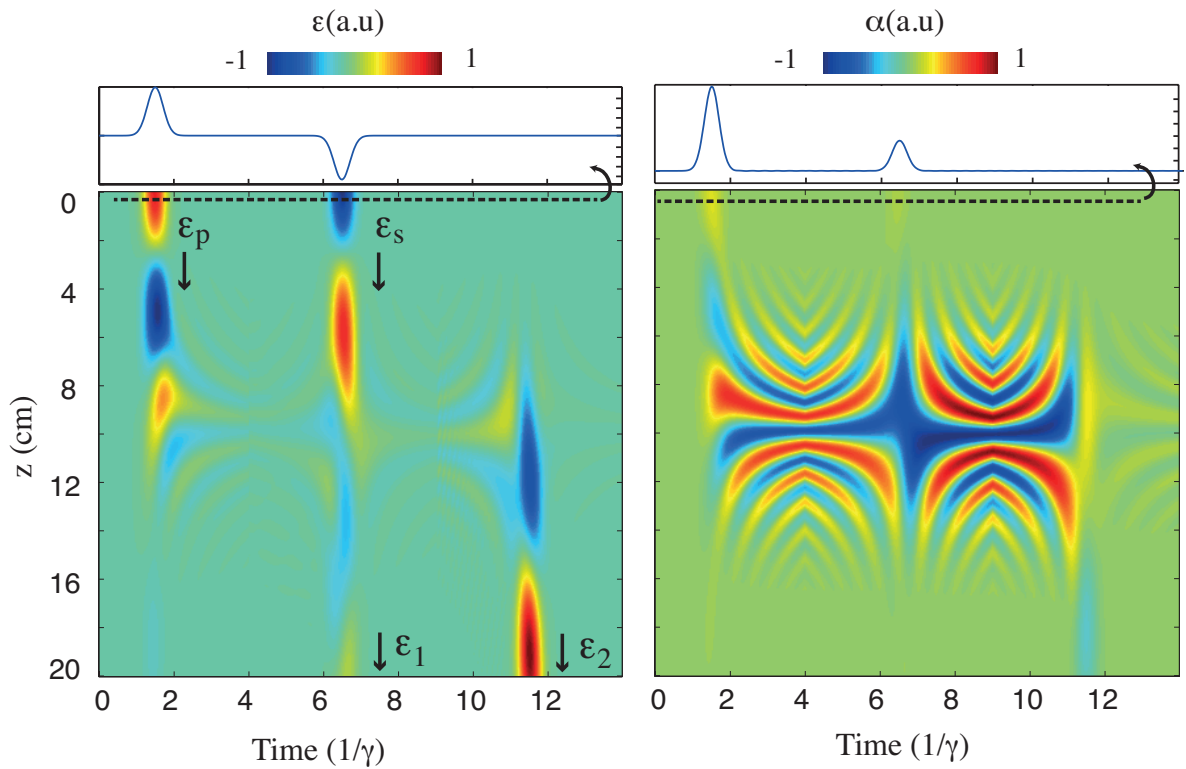}}
  \caption{The real parts of the electric field (left) and the atomic coherence (right) for a simulation of the time-domain interference experiment. The insets above the plots show the electric field and atomic coherence at the input of the ensemble.}
  \label{amp-zt}
\end{figure}

Figure \ref{amp-zt} shows the real components of the electric field and the atomic coherence to illustrate the interference for the same simulation parameters. The phase of the amplitude transmission and reflection coefficients in our beamsplitter analog can be extracted from the Maxwell equation, which gives $k\mathcal{E}(t,k)=N\frac{\Omega_{c}}{\Delta} \hat{\sigma}_{12}(t,k)$ \cite{Hetet:2008p7696}. The phase of the recalled probe field depends on both the phase of the coupling field, $\Omega_{c}$, and the phase of the atomic coherence, $\hat{\sigma}_{12}(t,k)$. This provides optical control over the relative phase between the two inputs to the interference. Note that when the normal mode crosses $k=0$ the relative phase of $\mathcal{E}(t,k)$ and $\hat{\sigma}_{12}(t,k)$ changes by $\pi$. This is analogous to the $\pi$ phase shift experienced on one side of an optical beamsplitter and is a requirement for the conservation of energy.

The experiment was performed using an ensemble of warm $^{87}$Rb atoms and a linear switchable varying magnetic field as described in Ref.~\cite{Hosseini:NComm:2011}. The coupling and probe fields are passed through acousto-optic modulators (AOMs) that allow us to create the required pulse sequences by driving them with appropriate RF signals. To generate the probe and steering pulses, the RF signals were created using separate, but phase locked, arbitrary waveform generators and were combined together before the AOM (details are in the supplementary material). In this manner, the frequency, phase and amplitude of the coupling, probe and steering fields can be independently controlled. The coupling field power used for maximum coupling between the optical and atomic modes was 330 mW and was adjusted to control the coupling. The probe and steering pulses have the smae frequency and were on the order of few $\mu$W. The coupling field, blue detuned by 3 GHz from the $S_{1/2}, F=2\rightarrow P_{1/2}, F^{\prime}=2$ transition, is Raman resonant with the probe and steering pulses which are blue detuned from the $S_{1/2}, F=1\rightarrow P_{1/2}, F^{\prime}=2$ transition.

We stored a 4 $\mu s$ probe pulse in the memory and recalled it after a storage time of $\tau_1 = 10 \mu s$. The steering pulse was injected just as the atomic coherence excited by the probe returned to $k=0$. We label the light detected at this time as $\mathcal{E}_1$, and integrate the detector signal over the pulse duration to obtain a value for the pulse energy. The polariton that remains in the atomic medium after the first recall is itself recalled after storage time $\tau_2 = 10 \mu s$. We detect it in the same manner as $\mathcal{E}_1$ and label it $\mathcal{E}_2$. Figure ~\ref{scheme}(a) shows the sequence of pulses that are stored, interfered and retrieved along with the coupling field intensity for each step.

\begin{figure}[!ht]
 \centerline{   \includegraphics[width=\columnwidth]{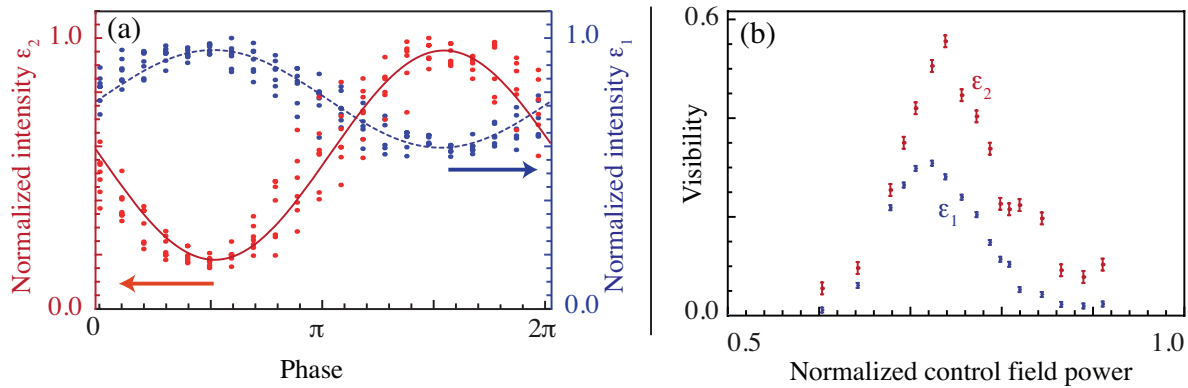}}
\caption{(a) Atom-light interference fringes at different times resulted from interaction of the steering pulse with echo generated from the probe pulse. The first arm of the interferometer which is in the optical mode leaves the memory(blue data) and the second arm is stored as an atomic coherence that is transformed back to the light field after re-switching the B-field (red data). The dashed blue and solid red lines are sinusoidal fits to the corresponding data. The red and blue data yield a fringe visibility of 68\% and 23\%, respectively. (b) Visibility of fringes for two pulses separated in time at the first (blue points) and second (red points) reading stage as a function the normalised coupling field power.}
 \label{tdbs}
\end{figure}

The energies of the recalled pulses, $\mathcal{E}_1$ and $\mathcal{E}_2$, were measured as a function of the relative phase of the probe and steering pulses. The phase of the atomic coherence depends on the relative phase of the coupling and probe fields. It is therefore possible to control the phase of the interference by scanning the phase of either the steering pulse or the corresponding coupling field. Fig.~\ref{tdbs}(a) shows interference fringes for $\mathcal{E}_1$ (blue, dashed line) and $\mathcal{E}_2$ (red, solid line) obtained by varying the phase of the coupling field corresponding to the steering pulse. This was accomplished by varying the phase of the RF signal that drives the coupling field AOM during the interference event relative to its phase during the storage of the probe pulse. For this data, the powers of $\mathcal{E}_p$ and $\mathcal{E}_s$ were equal and the coupling field power during the interference event was tuned to find the maximum fringe visibility on $\mathcal{E}_2$, which was found to be 68\%. The visibility of $\mathcal{E}_1$ echo, 23\%, is substantially lower due to the power mismatch of the steering pulse and the recalled atomic coherence required to optimise the interference in $\mathcal{E}_2$. The reflectivity corresponding to the recall of $\mathcal{E}_1$ is 37\%.

Control over the effective beamsplitter ratio is demonstrated in Fig. \ref{tdbs} (b). It can be seen that by varying the coupling field power the effective splitting ratio can be tuned to find a maximum in the interference. For this data, the power contained in the steering pulse was adjusted to provide good visibility for both  $\mathcal{E}_1$ and $\mathcal{E}_2$. It is interesting to note that for strong coupling fields, one optical pulse is written into memory while another is being recalled with little interference between the two, analogous to a high-reflectivity beamsplitter. For a weak coupling field, on the other hand, the effective beamsplitter becomes fully transmissive, again meaning no interference between the pulses as the steering pulse passes straight through without storage and the probe pulse remains trapped in the atomic coherence.

Now we consider the second experiment, in which the interference results from driving a single atomic coherence with multiple two-photon transitions as depicted in Fig.~\ref{scheme}(b). In this case, the probe and steering pulses are co-propagating and enter the medium simultaneously but are separated in frequency by more than the memory bandwidth. In the far-detuning and adiabatic regimes, this double-$\Lambda$ system is equivalent to a quasi-two-level system interacting with two fields of different Rabi frequency (see Fig. \ref{scheme} (b)). The interference between the two $\Lambda$ transitions will change the response of the medium to the probe and steering pulses. When they interfere destructively, the absorption of the probe and steering fields is suppressed and both pulses are transmitted through the medium. When the two $\Lambda$ transitions are in-phase, both pulses are coherently absorbed and can be recalled later on-demand.

As with our first experiment, the properties of the interference can be controlled through the coupling fields. The relative intensity and phase of the two coupling fields control the superposition of the probe and steering pulses that is transferred to the atomic coherence. This effect has been explored in EIT experiments \cite{Appel:2007p10361,Campbell:2009p3114}. Unlike EIT, however, the optical modes that are not coupled to the atomic coherence in the $\Lambda$-GEM scheme propagate through the atomic medium with little loss.

\begin{figure}[!t]
 \centerline{\includegraphics[width=0.7\columnwidth]{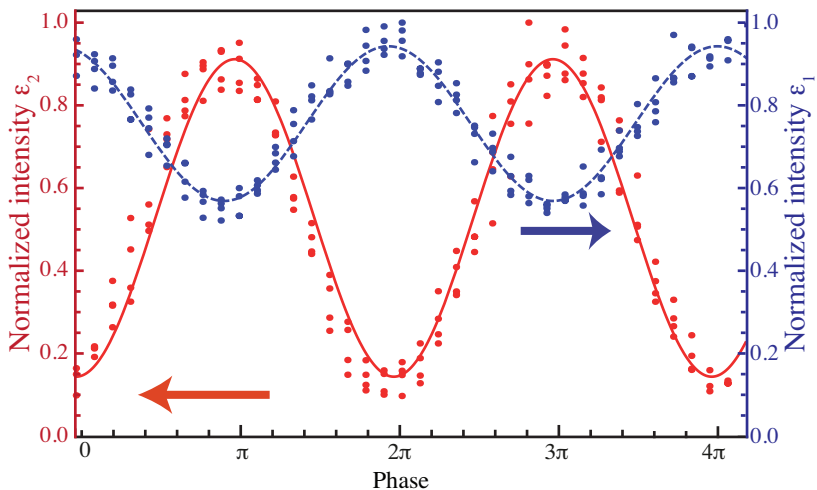}}
\caption{Interference fringes from $\mathcal{E}_1$ (blue) and $\mathcal{E}_2$ (red) that resulted from interference between double-$\Lambda$ transitions created from the probe and steering pulses of different frequency stored simultaneously in the memory. The dashed blue and solid red lines are sinusoidal fits to the corresponding data. The red and blue data yield a fringe visibility of 73\% and 25\%, respectively.}
 \label{fdbs}
\end{figure}

The frequency difference between the probe and steering fields was set to ~1 MHz, which was larger than the memory bandwidth of ~ 300 kHz to avoid overlap between two broadened Raman lines. Each of the probe and steering fields has a corresponding coupling field which is tuned to Raman resonance. The pulse length, 4 $\mu s$, was chosen to give a bandwidth slightly less than the memory bandwidth. Fig.~\ref{fdbs} shows the interference fringe obtained by varying the relative phase between the two Raman absorption lines. This was done by sweeping the phase of one of the coupling fields. The powers of the coupling fields are equal, 160 mW each, and remain constant throughout the storage and retrieval process. $\mathcal{E}_1$ is the portion of the probe and steering pulses that does not get stored in the memory and $\mathcal{E}_2$ is the portion that is retrieved from the memory after a 10 $\mu s$ storage period. The energies of $\mathcal{E}_1$ and $\mathcal{E}_2$ are measured by integrating the detector signal over the pulse period.

From an operational standpoint, this second experiment can be thought of as the frequency-domain counterpart to the first. While the first experiment demonstrated a beamsplitting operation between two pulses separated in time, the second demonstrates a beamsplitting operation between simultaneous pulses separated in frequency.

In both the time and frequency domain interference experiments we attribute the less-than-unity fringe visibility primarily to spatial and temporal mode mismatch between the probe polariton and the steering pulse. We believe that this is mainly due to the atomic motion and non-zero transverse magnetic field, which affects the echo signal for long storage times.  This can be justified by the larger visibility measured in the frequency-domain interference scheme, where interference occurs between pulses that simultaneously interact with the atomic coherence. During the storage time, atomic diffusion can change the spatial mode of the coherence and as a result, the echo signal will have a slightly different mode compared to the input signal. This effect is negligible for shorter storage times. The presence of a transverse magnetic field can induce an extra spatial frequency ($k_x$ and $k_y$) during storage. The transverse $k$ vector is imprinted to the echo signal at the readout diverting the output optical field slightly compare to the steering pulse.  In the beamsplitter analogy, this amounts to a poorly aligned interferometer. An inhomogeneous longitudinal magnetic field can alter the shape of the echo signal compared to its input leading to temporal mode mismatch. We anticipate, therefore, that the visibility could be improved by increasing the buffer gas pressure or using a cold atomic sample in order to increase the time of flight of the atoms and taking extra care with the magnetic environment to prevent pulse deflection and distortion. For the time-domain interference, numerical simulations (see supplementary material) reveal that, in the limit of large OD,  the interference visibility of the system can approach unity for both interfereometer outputs.

In summary, we have demonstrated interference effects between propagating optical fields and a collective atomic spin coherence. Fringe visibilities of 68\% and 73\% were observed for time-domain and frequency-domain interference schemes, respectively. These schemes may have relevance to manipulating optical quantum information. Unlike previous schemes, interference in a gradient echo memory could offer dynamic, optically addressable linear operations on optical qubits. These gates could operate on either time-bin or frequency multiplexed qubits or even a combination thereof. The time-delayed beamsplitter scheme can also be used for optical quantum state engineering \cite{Lvovsky:NPh:2010, Gisin:timbinInt:PRL:2000} and also for optimal Gaussian purification of coherent states from several imperfect copies \cite{FilipPRA:2005}. The ability to construct this type of interferometer is also of interest in building a coherent all-optical switch \cite{Hui:OSW:PRL:2005,Schmidt:OSW:2000}.

This research was conducted by the Australian Research Council Centre of Excellence for Quantum Computation and Communication Technology (project number CE110001027).

\bibliographystyle{unsrt}
\bibliography{bibs}

\begin{thebibliography}{10}

\bibitem{RevModPhys.70.1003}
K.~Bergmann, H.~Theuer, and B.~W. Shore.
\newblock Coherent population transfer among quantum states of atoms and
  molecules.
\newblock {\em Rev. Mod. Phys.}, 70:1003--1025, Jul 1998.

\bibitem{RevModPhys.77.633}
Michael Fleischhauer, Atac Imamoglu, and Jonathan~P. Marangos.
\newblock Electromagnetically induced transparency: Optics in coherent media.
\newblock {\em Rev. Mod. Phys.}, 77(2):633--673, Jul 2005.

\bibitem{Fleischhauer:2000p7451}
M.~Fleischhauer and M.~D. Lukin.
\newblock Dark-state polaritons in electromagnetically induced transparency.
\newblock {\em Phys. Rev. Lett.}, 84(22):5094--5097, May 2000.

\bibitem{Hartmann:echoint:OL:1993}
M.~Arend, E.~Bloch, and S.~R. Hartmann.
\newblock {\em Opt. Lett.}, 18(21):1789--1791, 1993.

\bibitem{Moiseev:TDSP:2001}
S.~A. Moiseev.
\newblock {\em Quant. Elect.}, 31:557--563, 2001.

\bibitem{Gisisn:IntEcho:PRL:2007}
M.~U. Staudt and et. al.
\newblock Interference of multimode photon echoes generated in spatially
  separated solid-state atomic ensembles.
\newblock {\em Phys. Rev. Lett.}, 99:173602, 2007.

\bibitem{Hetet:2008p5840}
G~H{\'e}tet, M~Hosseini, B.~M Sparkes, D~Oblak, Ping~Koy Lam, and Ben~C
  Buchler.
\newblock Photon echoes generated by reversing magnetic field gradients in a
  rubidium vapor.
\newblock {\em Opt. Lett.}, 33(20):2323--2325, 2008.

\bibitem{Hosseini:2009p8466}
Mahdi Hosseini, Ben~M Sparkes, G~Hetet, Jevon~J Longdell, Ping~Koy Lam, and
  Ben~C Buchler.
\newblock Coherent optical pulse sequencer for quantum applications.
\newblock {\em Nature}, 461(7261):241, Sep 2009.

\bibitem{Hosseini:NComm:2011}
M.~Hosseini, Ben~M. Sparkes, Geoff Campbell, Ping~K. Lam, and Ben~C. Buchler.
\newblock High efficiency coherent optical memory with warm rubidium vapour.
\newblock {\em Nat. Commun.}, 2(174), Feb 2011.

\bibitem{Hosseini:Nphys:2011}
M.~Hosseini, Geoff Campbell, Ben~M. Sparkes, Ping~K. Lam, and Ben~C. Buchler.
\newblock Unconditional room temperature quantum memory.
\newblock {\em Nat. Phys., DOI: 10.1038/NPHYS2021}, 2011.

\bibitem{Buchler:2010p11952}
B.~C Buchler, M~Hosseini, G~Hetet, B.~M Sparkes, and P.~K Lam.
\newblock Precision spectral manipulation of optical pulses using a coherent
  photon echo memory.
\newblock {\em Opt. Lett.}, 35(7):1091--1093, 2010.

\bibitem{Hetet:2008p7696}
G~Hetet, J.~J Longdell, M.~J Sellars, Ping~Koy Lam, and B.~C Buchler.
\newblock Multimodal properties and dynamics of gradient echo quantum memory.
\newblock {\em Phys. Rev. Lett.}, 101(20):203601--4, 2008.

\bibitem{Longdell:2008p8530}
J.~J Longdell, G~Hetet, Ping~Koy Lam, and M.~J Sellars.
\newblock Analytic treatment of controlled reversible inhomogeneous broadening
  quantum memories for light using two-level atoms.
\newblock {\em Phys. Rev. A}, 78(3):032337, 2008.

\bibitem{Appel:2007p10361}
J~Appel, E~Figueroa, and A~Lvovsky.
\newblock Adiabatic frequency conversion of optical information in atomic
  vapor.
\newblock {\em Opt. lett.}, 32(19):2771--2773, 2007.

\bibitem{Campbell:2009p3114}
Geoff Campbell, Anna Ordog, and A~I Lvovsky.
\newblock Multimode electromagnetically induced transparency on a single atomic
  line.
\newblock {\em New J. of Phys.}, 11:103021, 2009.

\bibitem{Lvovsky:NPh:2010}
Erwan Bimbard, Nitin Jain, Andrew MacRae, and A.~I. Lvovsky1.
\newblock Quantum-optical state engineering up to the two-photon level.
\newblock {\em Nat. Phot.}, 4:243--247, 2010.

\bibitem{Gisin:timbinInt:PRL:2000}
W.~Tittel, J.~Brendel, H.~Zbinden, and N.~Gisin.
\newblock Quantum cryptography using entangled photons in energy-time bell
  states.
\newblock {\em Phys. Rev. Lett.}, 84(20):4737--4740, May 2000.

\bibitem{FilipPRA:2005}
Ulrik~L. Andersen, Radim Filip, Jarom{\'\i}r Fiur{\'a}{\v s}ek, Vincent Josse,
  and Gerd Leuchs.
\newblock Experimental purification of coherent states.
\newblock {\em Phys. Rev. A}, 72:060301(R), 2005.

\bibitem{Hui:OSW:PRL:2005}
Jin-Hui Wu, Jin-Yue Gao, Ji-Hua Xu, L.~Silvestri, M.~Artoni, G.~C. La~Rocca,
  and F.~Bassani.
\newblock Ultrafast all optical switching via tunable fano interference.
\newblock {\em Phys. Rev. Lett.}, 95(5):057401, Jul 2005.

\bibitem{Schmidt:OSW:2000}
H.~Schmidt and R.~J. Ram.
\newblock All-optical wavelength converter and switch based on
  electromagnetically induced transparency.
\newblock {\em App. Phys. Lett.}, 76:3173--3175, 2000.

\end{thebibliography}

\end{document}